# Verification of Two-Dimensional Monte Carlo Ray-Trace Methodology in Radiation Heat Transfer Analysis

Mehran Yarahmadi and J. Robert Mahan
Virginia Polytechnic Institute and State University, mehrany@vt.edu, jrmahan@vt.edu

*Abstract* - **Despite the frequent appearance in the radiation heat transfer literature of articles describing Monte Carlo ray-trace (MCRT) applications to two-dimensional enclosures, no formal verification may be found of the method commonly used to determine the directional distribution of diffuse emission and reflection when estimating two-dimensional radiation distribution factors. Considered are two methods for determining the direction cosines in this situation. The results are shown to be in agreement with those obtained in the limiting case of a three-dimensional enclosure as one of its dimensions is increased.**

## 1. INTRODUCTION

Radiation is the dominant mode of heat transfer in many applications of practical engineering interest. These include situations, such as instrumentation, cryogenics, solar energy utilization, and certain space applications, where other modes of heat transfer have been suppressed or eliminated; and high-temperature processes, such as those associated with combustion and detonation.

The usual starting point in radiation heat transfer analysis is to define an enclosure whose walls are typically subdivided into surface elements of size depending on the desired spatial resolution. It is sometimes convenient to treat the enclosure as being two-dimensional. This occurs when one of the three dimensions is long compared to the other two, as in the case of ducts and certain industrial process lines.

Because of their relative simplicity, two-dimensional enclosures have been widely used in the radiation heat transfer literature to establish epistemology regarded as independent of dimensionality. For example, Chang et al. [1] have investigated the effect of radiation on combined heat transfer with convection or conduction in a participating medium in a two-dimensional enclosure. Ramankutty et al. [2] demonstrate a modified discrete ordinates solution of radiative transfer in two-dimensional rectangular enclosures. Ismail and Salinas [3] study the application of a multidimensional scheme using the discrete ordinate method in a two-dimensional enclosure with diffusely emitting and reflecting walls. Hayasaka et al. [4] consider the radiative heat ray method in a two-dimensional model. Jinbo et al. [5] investigate the radiative heat fluxes and temperatures under the assumption of isotropic scattering in a two-dimensional stationary rectangular configuration. Two-dimensional systems have also been investigated for numerical studies of radiation of water droplet systems [6-11].

Many investigators have used two-dimensional enclosures for inverse boundary design in radiation heat transfer. Li [12] considers the inverse problem of an unknown source term in a two-dimensional rectangular medium with transparent boundaries. Sarvari et al. [13, 14] present an inverse analysis for finding the heat source distribution in an irregular enclosure to produce both desired temperature and heat flux profiles over the design surface of an irregular two-dimensional enclosure with participating media. Tito et al. [15] consider inverse radiative transfer problems in two-dimensional rectangular enclosures containing heterogeneous isotropic scattering or linear anisotropic scattering participating media. Daun et al. [16, 17], use optimization methods for finding the heater settings that provide spatially uniform transient heating within a two-dimensional radiant enclosure. The variable metric method is utilized by Kowsary et al. [18] to investigate the radiative boundary design problem in a two-dimensional furnace filled with an absorbing, emitting and scattering gas. The conjugate gradient method has been applied to inverse boundary design problems in an irregular two-dimensional enclosure with participating media by Pourshaghaghy et al. [19]. Mehraban et al. [20] present an inverse radiation design problem for finding the transient heater settings to produce the transient conditions over products in two-dimensional radiant furnaces. Salinas [21] present an optimization analysis for temperature field estimation in a two-dimensional gray medium. Bayat et al. [22] use the conjugate gradient method to investigate an optimization procedure to determine the heater powers of a radiant enclosure to achieve a uniform heat flux distribution over a diffuse-spectral temperature-specified design surface in a two-dimensional radiant furnace. Amiri et al. [23] employ an inverse analysis to estimate the required input on the heater surface that produces the desired temperature and heat flux distribution over the design surface of a two-dimensional enclosure.

The inverse boundary design problem for combined radiation convection/conduction heat transfer in two-dimensional enclosures is also studied. An optimization technique has been applied to the design of two-dimensional heat transfer systems in which both conduction and radiation are important [24]. Kim et al. [25] investigate an inverse problem based on the finite volume method for conduction and radiation in a two-dimensional cylindrical enclosure. Mossi et al. [26] report an inverse boundary design problem involving radiation and convection in a two-dimensional cavity. Moghadassian et al. [27] investigate the inverse boundary design problem in combined natural convection radiation heat transfer with the presence of a participating medium in a square two-dimensional square.

The Monte Carlo ray-trace (MCRT) method is a statistical solution technique in which energy bundles are traced as they are emitted, scattered, and absorbed within an enclosure. The method produces very accurate solutions within limits of statistical accuracy, which can be estimated. In addition to the applicability of the MCRT method in complex geometries, it is also a very flexible method and benefits from a straightforward formulation that enables easier handling of further complexities such specularly reflecting surfaces, effects of participating medium and scattering [28, 29]. Furthermore, compared to other methods such as the finite-volume method (FVM), the discrete ordinates method (DOM), the discrete transfer method (DTM), and the finite element method (FEM), the MCRT method avoids the ray effect and false scattering [30].

Numerous two-dimensional studies have been based on the Monte Carlo ray-trace method. Oguma and Howell [31] investigate the solution of two-dimensional blackbody inverse radiation problems by the inverse Monte Carlo method. Erturk [32] considers a two-dimensional inverse design approach using a combination of MCRT and regularization methods. Baek et al. [33] consider a combination of the Monte-Carlo and finite-volume methods (CMCFVM) for solving radiative heat transfer in absorbing, emitting, and isotropically scattering medium with an isolated boundary heat source in a two-dimensional irregular geometry. Safavinejad et al. [34] use a micro-genetic algorithm to solve the inverse boundary design problem in two-dimensional radiant enclosures with absorbing–emitting media. They use the Monte Carlo method (MCM) to solve the equation of radiative transfer. In a second contribution, they used the same method to optimize the number and location of the heaters in two-dimensional radiant enclosures composed of specular and diffuse surfaces [35]. Mosavati et al. [36] apply the MCRT method in a two-dimensional enclosure for calculating distribution factors used in an inverse design problem. By employing the backward Monte Carlo method for computing the distribution factors, they also solved the boundary inverse design in a step-like two-dimensional enclosure with gray walls and a transparent medium with combined radiating-free convection [37]. In a recent study, Mulford et al. [38] apply two-dimensional Monte Carlo ray-tracing to calculate the apparent absorptivity of a diffusely-irradiated V-groove and the apparent absorptivity of a fully illuminated cavity subject to collimated irradiation.

In contrast to heat conduction and other boundary value problems, the Monte Carlo ray-trace method does not involve solution of differential equations but rather is based on the rules of statistics and geometrical optics. The rules governing diffuse emission and reflection in a three-dimensional enclosure are well established [28, 29]; however, it seems that the rules governing a two-dimensional ray-trace, while perhaps intuitively obvious and certainly widely used, have yet to be rigorously established in the literature. In the current contribution, the frequent use of the two-dimensional approximation for elongated enclosures is critically examined. Specifically, two candidate methods for determining the direction of diffuse emission and reflection are investigated. Finally, the results obtained using these two methods are compared with those obtained for an equivalent three-dimensional enclosure in the limit as its long dimension is extended.

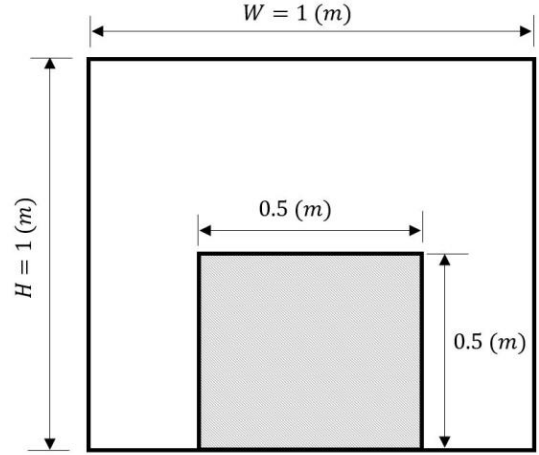

**Figure 1**. Cross-section of the two-dimensional furnace.

**2. Problem Description**

Consider the radiation problem, illustrated in Fig. 1, involving a two-dimensional furnace whose lower surface has a step-like geometry. All surfaces are considered to be gray and diffuse with an emissivity of 0.8, and the interior medium is assumed to be non-participating. This enclosure has been selected because it is a benchmark geometry in the literature [16, 18, 19, 23, 35-37]. The problem is to first calculate the distribution factor matrix for this enclosure using the two-dimensional MCRT method, and then compare the results with those describing the equivalent elongated three-dimensional enclosure that it is intended to represent. The total number of surface elements for this problem is selected to be 40. Fig. 2 shows the equivalent three-dimensional enclosure. Note that for comparison of the results of two-dimensional and three-dimensional enclosures, the surface elements in three-dimensional enclosures are chosen to be long strips, as indicated in Fig. 2.

The radiation distribution factors are most easily determined using the Monte Carlo ray-trace method, as detailed by Mahan [29]. Briefly, the steps for obtaining these factors for a diffuse-gray enclosure are:

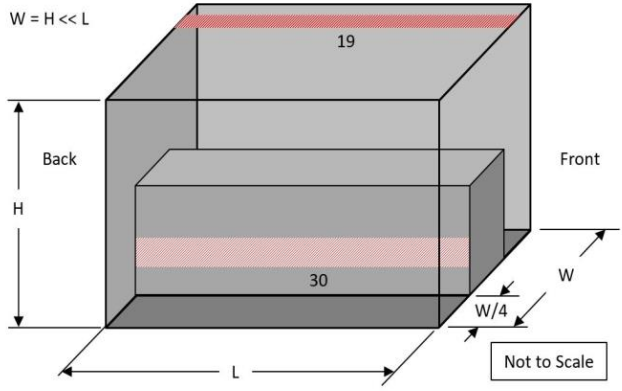

**Figure 2**. Isometric view of the long three-dimensional oven. ($W = H = 1\ (m)$)

(1) The randomly distributed points from which discrete rays are launched from any given surface element are selected based on the values of two random numbers.

(2) The directions of diffuse emission are determined by drawing two additional random numbers from which are calculated azimuth and zenith angles.

(3) The intersection point of the emitted rays with the enclosure interior surface is determined.

(4) Whether the ray is reflected or absorbed by the surface intersected by the ray is determined by drawing a fifth random number and comparing its value to the surface absorptivity. If the random number is less than the absorptivity, the ray is absorbed and its history is terminated. In this case the counter keeping track of the rays absorbed by the intersected surface element is incremented by one.

(5) If the ray is reflected, the diffuse direction is determined by returning to Step 2 and repeating the procedure until the ray is finally absorbed by one of the surfaces of the enclosure. The ratio of the number of rays absorbed by surface $j$ to those emitted from surface $i$ is an estimator of the radiation distribution factor $D_{ij}$, assuming a sufficient number of rays have been emitted.

The uncertainty of the results obtained using the MCRT method may result from the measuring errors of parameters such as temperature and emissivity [39, 40]. A detailed analysis of the uncertainty in the MCRT environment is available [41].

### 3. FINDING THE MINIMUM LENGTH OF THE THREE-DIMENSIONAL OVEN

A numerical experiment has been carried out using a standard three-dimensional Monte Carlo ray-trace for oven lengths starting from 1 m and increasing in steps of 1 m. Figure 3 is a plot of the fraction, $D_{ij}$, of the energy emitted from two surface elements ($i$ = 19 or 30), indicated as strips in Fig. 2, that is absorbed on either the front or back surface ($j$ = front or back) of the three-dimensional enclosure due to emission.

When the length $L$ is 100 m, the fraction of energy emitted by elements 19 and 30 absorbed by the front or back surfaces is only 0.17 percent and 0.10 percent, respectively, of the emitted energy. Therefore, it is reasonable to assume that the three-dimensional oven with a length of $L$=100 m can be considered a two-dimensional enclosure.

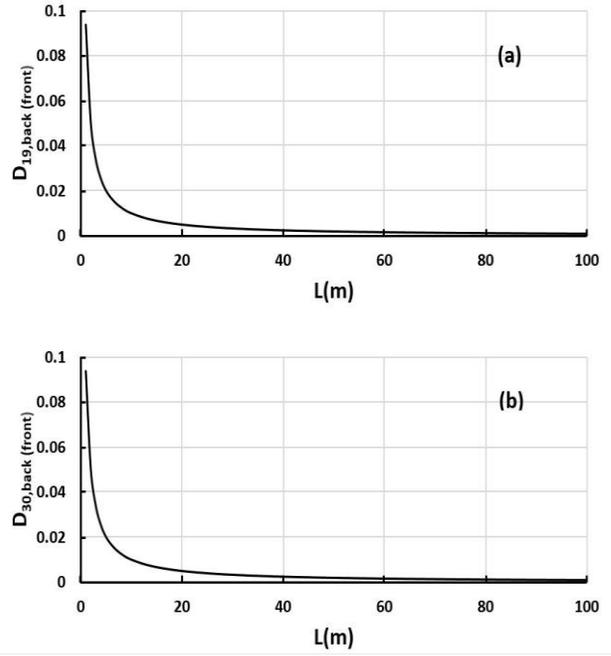

**Figure 3**. Distribution factor value from two selected three-dimensional elements **(a)** 19 and **(b)** 30 to the back (front) surfaces.

### 4. TWO-DIMENSIONAL MCRT METHOD

The only difference between the two-dimensional MCRT method and the three-dimensional method is the algorithm for computing the direction of diffuse emission and reflection. For three-dimensional analysis, the direction cosines $L$, $M$, and $N$ are determined as

$$L = n_x\cos\theta + t_{1,x}\sin\theta\cos\phi + t_{2,x}\sin\theta\sin\phi, \quad (1)$$
$$M = n_y\cos\theta + t_{1,y}\sin\theta\cos\phi + t_{2,y}\sin\theta\sin\phi, \quad (2)$$

and

$$N = n_z\cos\theta + t_{1,z}\sin\theta\cos\phi + t_{2,z}\sin\theta\sin\phi, \quad (3)$$

where $n$, $t_1$, and $t_2$ are the unit normal, first unit tangent, and second unit tangent vectors of each surface element of emission or reflection, and $\theta$ and $\phi$ are zenith and azimuth angles measured with respect to unit normal and tangent vectors. These two angles are randomly determined by

$$\theta = \sin^{-1}\left[\sqrt{R_\theta}\right] \quad and \quad \phi = 2\pi R_\phi, \quad (4)$$

where $R_\theta$ and $R_\phi$ are random numbers uniformly distributed between zero and unity.

In three-dimensional geometries we have three direction cosines, while in the two-dimensional analysis we have only two, $L$ and $M$, with respect to the global x and y axes, respectively. Two methods are investigated for finding the direction cosines of emission in two-dimensional geometries.

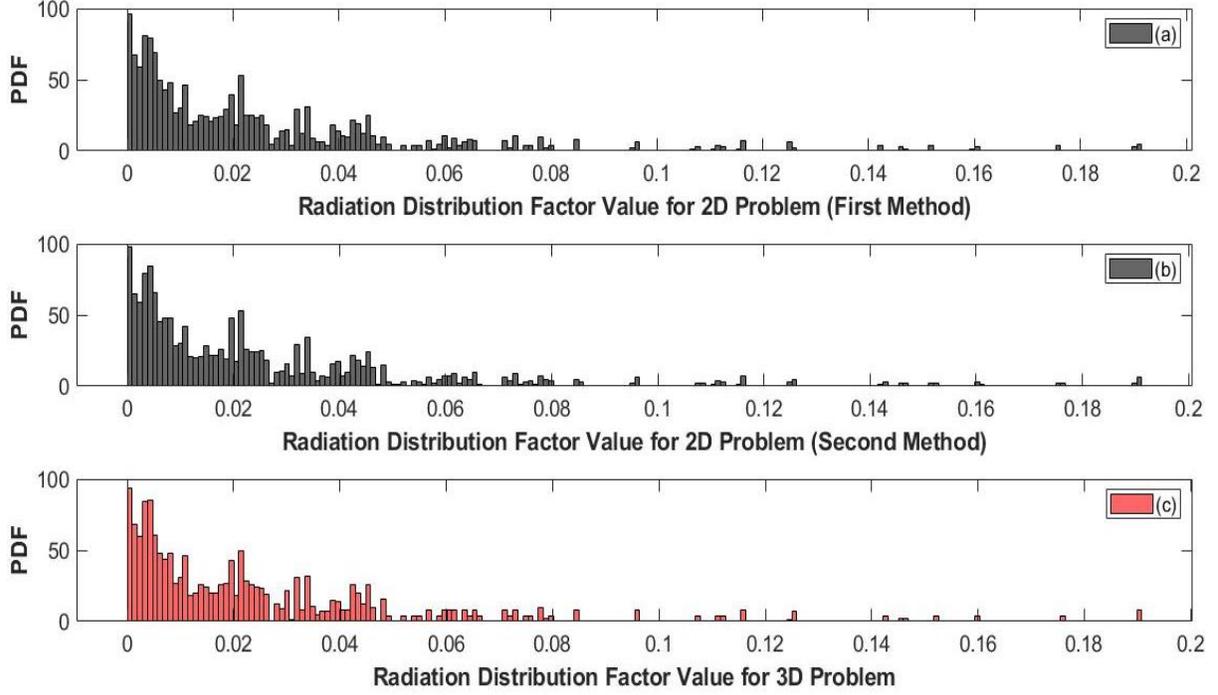

**Figure 4**. Probability density functions (PDFs) for radiation distribution factors for (a) two-dimensional problem using first method the (b) two-dimensional problem using the second method (c) three-dimensional problem with $L/H = L/W = 100$.

**First method**. This method is based on the same logic as for three-dimensional analysis, and uses two random numbers. Here, z-axis direction cosine, $N$, from Eq. (3) is forced to be zero, and the values obtained from Eqs. (1) and (2) are used for the other two direction cosines. Since the emission or reflection vector must be a unit vector, we normalize the values of $L$ and $M$ from Eqs. (1) and (2), obtaining

$$\boldsymbol{V} = \left[\frac{L}{\sqrt{L^2 + M^2}}, \frac{M}{\sqrt{L^2 + M^2}}\right]. \quad (5)$$

**Second method**. This method uses only one random number. Here, the angle $\alpha$ with the x-axis is randomly determined as

$$\alpha = 2\sin^{-1}[\sqrt{R_\alpha}], \quad (6)$$

where $R_\alpha$ is again a random number uniformly distributed between zero and unity. Then, for the direction cosines, we have

$$L = n_y \cos\theta + t_y \sin\theta \quad (7)$$

and

$$M = n_x \cos\theta + t_x \sin\theta. \quad (8)$$

## 5. RESULTS AND DISCUSSION

Radiation distribution factors are computed using a windows application [42] based on the MCRT method to compute the radiation distribution factors among any number of surface elements making up any two-dimensional diffuse gray enclosure. As shown by Yarahmadi et al. [41], an effective way to present the results for radiation distribution values is as a probability density function (PDF). Figure 4 shows the distribution factor values obtained using the two-dimensional MCRT analysis mentioned above compared with the results for the three-dimensional elongated enclosure discussed in Section 3. It is clear that both methods of two-dimensional analysis provide distribution factor matrices that are virtually identical to the results obtained using the three-dimensional MCRT method. The average difference between the two-dimensional and three-dimensional analysis is calculated based on

$$difference = \left\langle \frac{\left|D_{ij}^{3D} - D_{ij}^{2D}\right|}{\frac{1}{2}(D_{ij}^{3D} + D_{ij}^{2D})} \right\rangle, \quad i,j = 1,..,n \quad (9)$$

The results obtained using Method 1 are only 0.99 percent different from those obtained using the three-dimensional analysis, while this difference is 1.01 percent using Method 2. We can see that Method 1 is slightly more accurate than Method 2, however, Method 2 is faster because

**4**

it only requires a single random number to describe the direction of the emission or reflection.

## 6. CONCLUSION

We conclude that either of two methods, designated here as Method 1 and Method 2, can be reliably used to compute the radiation distribution factors for a two-dimensional enclosure.

### ACKNOWLEDGMENT

The authors gratefully acknowledge NASA's Langley Research Center for its financial support for this effort under NASA Contract NNL16AA05C with Science Systems and Applications, Inc., and Subcontract No. 21606-16-036, Task Assignment M.001C (CERES) with Virginia Tech.

**Uncategorized References**

1. Chang, L.C., K.T. Yang, and J.R. Lloyd, *Radiation-Natural Convection Interactions in Two-Dimensional Complex Enclosures.* Journal of Heat Transfer, 1983. **105**(1): p. 89-95.
2. Ramankutty, M.A. and A.L. Crosbie, *Modified discrete ordinates solution of radiative transfer in two-dimensional rectangular enclosures.* Journal of Quantitative Spectroscopy and Radiative Transfer, 1997. **57**(1): p. 107-140.
3. Ismail, K. and C. Salinas, *Application of multidimensional scheme and the discrete ordinate method to radiative heat transfer in a two-dimensional enclosure with diffusely emitting and reflecting boundary walls*. Vol. 88. 2004. 407-422.
4. H. Hayasaka, K.K., H. Taniguchi, I. Nakamachi, T. Omori, T. Katayama, *Radiative Heat Transfer Analysis by Radiation Heat Ray Method.* Japan Society of Mechanical Engineering, 1986. **52**: p. 1734-1740.
5. Jinbo, H., R. Liming, and T. Heping, *Effect of anisotropic scattering on radiative heat transfer in two-dimensional rectangular media.* Journal of Quantitative Spectroscopy and Radiative Transfer, 2003. **78**(2): p. 151-161.
6. Berour, N., et al., *Radiative and conductive heat transfer in a nongrey semitransparent medium. Application to fire protection curtains.* Journal of Quantitative Spectroscopy and Radiative Transfer, 2004. **86**(1): p. 9-30.
7. Collin, A., et al., *On radiative transfer in water spray curtains using the discrete ordinates method.* Journal of Quantitative Spectroscopy and Radiative Transfer, 2005. **92**(1): p. 85-110.
8. Consalvi, J.-L., B. Porterie, and J. C. Loraud, *On the Use of Gray Assumption for Modeling Thermal Radiation Through Water Sprays*. Vol. 44. 2003. 505-519.
9. Coppalle, A., D. Nedelka, and B. Bauer, *Fire protection: Water curtains.* Fire Safety Journal, 1993. **20**(3): p. 241-255.
10. Ravigururajan, T.S. and M.R. Beltran, *A model for attenuation of fire radiation through water droplets.* Fire Safety Journal, 1989. **15**(2): p. 171-181.
11. W. Yang, T.P., H. Ladouceur, R. Kee, *The interaction of thermal radiation and water mist in fire suppression.* Fire Safety Journal, 2004. **39**: p. 41-66.
12. Li, H.Y., *Inverse Radiation Problem in Two-Dimensional Rectangular Media.* Journal of Thermophysics and Heat Transfer, 1997. **11**(4): p. 556-561.
13. Hosseini Sarvari, S.M. and S. Mansouri, *Inverse design for radiative heat source in two-dimensional participating media*. Vol. Part B. 2004. 283-300.
14. Hosseini Sarvari, S.M., S.H. Mansouri, and J.R. Howell, *Inverse Boundary Design Radiation Problem in Absorbing-Emitting Media with Irregular Geometry.* Numerical Heat Transfer, Part A: Applications, 2003. **43**(6): p. 565-584.
15. Berrocal Tito, M.J., et al., *Inverse radiative transfer problems in two-dimensional participating media.* Inverse Problems in Science and Engineering, 2004. **12**(1): p. 103-121.
16. Daun, K.J. and J.R. Howell, *Inverse design methods for radiative transfer systems.* Journal of Quantitative Spectroscopy and Radiative Transfer, 2005. **93**(1): p. 43-60.
17. Daun, K.J.H., J. R.; Morton, D. P., *Optimization of Transient Heater Settings to Provide Spatially Uniform Heating in Manufacturing Processes involving Radiant Heating.* Numerical Heat Transfer: Part A: Applications, 2004. **46**(7): p. 651-667.
18. Kowsary, F., K. Pooladvand, and A. Pourshaghaghy, *Regularized variable metric method versus the conjugate gradient method in solution of radiative boundary design problem.* Journal of Quantitative Spectroscopy and Radiative Transfer, 2007. **108**(2): p. 277-294.
19. Pourshaghaghy, A., et al., *An inverse radiation boundary design problem for an enclosure filled with an emitting, absorbing, and scattering media.* International Communications in Heat and Mass Transfer, 2006. **33**(3): p. 381-390.
20. S. Mehraban, S.M.H.S., S. Farahat. *A Quasi-Steady Method for Inverse Design and Control of a Two-Dimensional Radiant Oven in Transient State*. in *ICHMT Int. Symposium on Advances in Computational Heat Transfer*. 2008. Marrakesh, Morocco
21. Salinas, C.T., *Inverse radiation analysis in two-dimensional gray media using the discrete ordinates method with a multidimensional scheme.* International Journal of Thermal Sciences, 2010. **49**(2): p. 302-310.
22. Bayat, N., S. Mehraban, and S.M.H. Sarvari, *Inverse boundary design of a radiant furnace with diffuse-spectral design surface.* International Communications in Heat and Mass Transfer, 2010. **37**(1): p. 103-110.
23. Amiri, H., S.H. Mansouri, and P.J. Coelho, *Inverse Optimal Design of Radiant Enclosures With Participating Media: A Parametric Study.* Heat Transfer Engineering, 2013. **34**(4): p. 288-302.
24. Hosseini Sarvari, S.M., J.R. Howell, and S.H. Mansouri, *Inverse Boundary Design Conduction-Radiation Problem in Irregular Two-Dimensional Domains.* Numerical Heat Transfer, Part B: Fundamentals, 2003. **44**(3): p. 209-224.
25. Kim, K.W., et al., *Estimation of emissivities in a two-dimensional irregular geometry by inverse radiation analysis using hybrid genetic algorithm.* Journal of Quantitative Spectroscopy and Radiative Transfer, 2004. **87**(1): p. 1-14.
26. Mossi, A.C., et al., *Inverse design involving combined radiative and turbulent convective heat transfer.* International Journal of Heat and Mass Transfer, 2008. **51**(11): p. 3217-3226.
27. Moghadassian, B. and F. Kowsary, *Inverse boundary design problem of natural convection–radiation in a square enclosure.* International Journal of Thermal Sciences, 2014. **75**: p. 116-126.
28. Mahan, J.R., *Radiation Heat Transfer: A Statistical Approach*. 2002, New York, NY: John Wiley & Sons.
29. Mahan, J.R., *The Monte Carlo Ray-Trace Method in Radiation Heat Transfer and Applied Optics*. 2018, New York, NY: John Wiley & Sons.
30. Tan, H.-P., H.-C. Zhang, and B. Zhen, *Estimation of Ray Effect and False Scattering in Approximate Solution Method for Thermal Radiative Transfer Equation.* Numerical Heat Transfer, Part A: Applications, 2004. **46**(8): p. 807-829.
31. M. Oguma, J.R.H. *Solution of Two-Dimensional Blackbody Inverse Radiation Problems by Inverse Monte Carlo Method*. in *ASME/JSME Joint Thermal Engineering*. 1995. Maui, Hawaii
32. Erturk, H., Ezekoye, O. A., and Howell, J. R. *Inverse Solution of Radiative Transfer in Two Dimensional Irregularly Shaped Enclosures*. in *ASME International Mechanical Engineering Congress and Exhibition*. 2000. Orlando, FL
33. Baek, S.W., D.Y. Byun, and S.J. Kang, *The combined Monte-Carlo and finite-volume method for radiation in a two-dimensional irregular geometry.* International Journal of Heat and Mass Transfer, 2000. **43**(13): p. 2337-2344.
34. Safavinejad, A., Mansouri, S. H., and Hosseini Sarvari, S. M, *Inverse Boundary Design of 2-D Radiant Enclosures with Absorbing-Emitting Media Using Micro-Genetic Algorithm.* IASME Transaction, 2004. **46**: p. 651-667.
35. Safavinejad, A., et al., *Optimal number and location of heaters in 2-D radiant enclosures composed of specular and diffuse*

surfaces using micro-genetic algorithm. Applied Thermal Engineering, 2009. **29**(5): p. 1075-1085.
36. Mosavati, M., F. Kowsary, and B. Mosavati, *A Novel, Noniterative Inverse Boundary Design Regularized Solution Technique Using the Backward Monte Carlo Method.* Journal of Heat Transfer, 2013. **135**(4): p. 042701-042701-7.
37. Mosavati, B., M. Mosavati, and F. Kowsary, *Solution of radiative inverse boundary design problem in a combined radiating-free convecting furnace.* International Communications in Heat and Mass Transfer, 2013. **45**: p. 130-136.
38. Mulford, R.B., et al., *Total hemispherical apparent radiative properties of the infinite V-groove with specular reflection.* International Journal of Heat and Mass Transfer, 2018. **124**: p. 168-176.
39. Moazami Goudarzi, H., M. Yarahmadi, and M.B. Shafii, *Design and construction of a two-phase fluid piston engine based on the structure of fluidyne.* Energy, 2017. **127**: p. 660-670.
40. Yarahmadi, M., H. Moazami Goudarzi, and M.B. Shafii, *Experimental investigation into laminar forced convective heat transfer of ferrofluids under constant and oscillating magnetic field with different magnetic field arrangements and oscillation modes.* Experimental Thermal and Fluid Science, 2015. **68**: p. 601-611.
41. Yarahmadi, M., J.R. Mahan, and K.J. Priestley, *Uncertainty Analysis and Experimental Design in the Monte Carlo Ray-Trace Environment.* Journal of Heat Transfer, 2018.
42. Yarahmadi, M., *An Engine for Computing Radiation Distribution Factors for Two-Dimensional Enclosures.* Zenodo, 2018. http://doi.org/10.5281/zenodo.1407154.